\documentclass[prb,twocolumn,showpacs,preprintnumbers,amsmath,amssymb]{revtex4}
\usepackage{setspace}
\usepackage{subfigure}
\usepackage{graphicx, epsfig}
\usepackage{dcolumn}
\def\be{\begin{equation}}
\def\ee{\end{equation}}
\def\bea{\begin{eqnarray}}
\def\eea{\end{eqnarray}}
\begin{document}

\title{Hydrogen dynamics and light-induced structural changes in hydrogenated amorphous silicon}

\author{T. A. Abtew}
\email{abtew@phy.ohiou.edu}

\author{D. A. Drabold}
\email{drabold@ohio.edu}

\affiliation{Department of Physics and Astronomy, Ohio University, 
Athens, OH 45701}


\date{\today}
\begin{abstract}

We use accurate first principles methods to study the network dynamics of hydrogenated amorphous silicon, including the motion of hydrogen. In addition to studies of atomic dynamics in the electronic ground state, we also adopt a simple procedure to track the H dynamics in light-excited states. Consistent with recent experiments and computer simulations, we find that dihydride structures are formed for dynamics in the light-excited states, and we give explicit examples of pathways to these states. Our simulations appear to be consistent with aspects of the Staebler-Wronski effect, such as the light-induced creation of well separated dangling bonds. 
\end{abstract}
 
\maketitle

\section{INTRODUCTION}

A variety of experiments and many theoretical studies of light-induced structural changes in hydrogenated amorphous silicon (a-Si:H) reveal the complexity of photoresponse in these materials. Light-induced changes in photoconductivity and defect formation~\cite{Staebler,stutzmann,han}, enhanced hydrogen diffusion~\cite{jackson,santos,Branz,branz,cheong}, and creation of preferential proton separation~\cite{Taylor} are among the phenomena observed experimentally in a-Si:H. A special feature of these materials is the Jekyll-Hyde behavior of hydrogen as a defect passivator enabling the practical utilization of the material  but also as a culprit in light-induced defect creation \cite{zellama,kemp,tuttle,pantelides}. A variety of experiments implicate H and its dynamics as being important player to the SWE\cite{street}.

There are various models proposed to explain the light-induced metastability. Chang {\it et al.} \cite{Chang,chang2,chang3} suggested that the dissociation of a two hydrogen interstitial complex, (H$_{2}^{\ast}$), into separate and more mobile H atoms, caused by carriers localized on the H$_{2}^{\ast}$, is a mechanism for the metastable phenomena. In the hydrogen flip model, Biswas {\it et al.} demonstrated a higher energy metastable state is formed when H is flipped to the backside of the Si-H bond at a monohydride sites~\cite{biswas}. In the hydrogen collision model proposed by Branz \cite{Branz}, the recombination-induced emission of H from Si-H bond creates  mobile H and dangling bonds, and these newly created dangling bonds become metastable when two mobile H atoms collide to form a metastable complex containing two Si-H bonds. The two-phase model of Zafar and Schiff~\cite{zafar1, zafar2}, explained thermal stability data, exploited the concept of paired hydrogen, and later merged with Branz model and invoked dihydride bonding~\cite{kopidakis}. Though they have different detailed mechanisms for the formation of the structures, these models share the same notion that the diffusion of H and H pair formation plays a key role in light-induced metastability. From {\it ab initio} simulation in a photo-excited state, Fedders, Fu and Drabold showed that changes of charge of well-localized defect states could induce defect creation~\cite{fedders}. Our work can be regarded as a natural extension of Ref. \onlinecite{fedders} to the case of a hydrogenated system, and with more accurate techniques than those available at the time of the original study.

Our work has been particularly driven by the experimental work of Su {\it et al.} \cite{Taylor}, who performed proton NMR experiments in a-Si:H and found preferential creation of H-H distance of 2.3$\pm$0.2 \AA.  We have shown that SiH$_2$ configurations in the solid state are consistent with these observations \cite{Tesdad2}. More recently, an NMR study by Bobela {\it et al} \cite{bobela}, indicated a shorter proton-proton distance in a sample of a-Si:H with a somewhat higher defect density.

In a recent paper\cite{tesdad3}, we briefly reported the hydrogen dynamics and light-induced formation of a SiH$_2$ structure for a small (71 atom) model. In this article, we provide detailed results of simulation on electronic properties, vibrational properties and mechanism of hydrogen diffusion in both the electronic ground state and light excited state in a-Si:H using two different supercell models: a larger 223 atom a-Si:H model (Model-I) and a 71 atom a-Si:H model (Model-II). We start with small, but topologically credible models of a-Si:H and use a sufficiently accurate method to simulate network dynamics, including the diffusive motion of hydrogen. In a photo-excited MD simulation, we find enhanced H motion and the preferential formation of a paired-H final states, a significant confirmation of aspects of some current influential models~\cite{Branz}. Furthermore, we obtain final states with additional dangling bond defects, well separated from each other, again in agreement with key experiments~\cite{isoya}. Our simulations suffer from limitations: small cells and certainly a limited sampling of the possible motions of H, limited simulation times and approximations of various forms described below. Nevertheless, we believe that studies of this type offer significant promise as an ``unbiased" means to discover the importance of H motion, its topological and electronic consequences, and provide another needed piece to a complex puzzle.

The rest of the paper is organized as follows. In Sec.~\ref{secII} we discuss the approximations used in the {\it ab initio} local basis code SIESTA \cite{ordejon,sanchez,soler}. describe the procedure for generating the models and also discuss the methods used to simulate both the electronic ground state and the light excited state.  In Sec.~\ref{secIII} we present detailed discussions of the molecular dynamics calculations of Hydrogen dynamics and its consequences both in the electronic ground state and in the presence of light excitation. The change in the electronic structures and vibrational modes are explained in detail. We present conclusions in Sec.~\ref{secIV}.

\section{METHODOLOGY}
\label{secII}
\subsection{Total energies, electronic structure and dynamical simulation}

Simulation of complex a-Si:H models require accurate interatomic interactions (in particular, H energetics is highly delicate in a-Si:H)~\cite{vandewalle-n,attafynn1-n}. Therefore, our density functional simulations were performed within the generalized gradient approximation\cite{perdew} (GGA) or the local density approximation (LDA) using the first principles code SIESTA \cite{ordejon,sanchez,soler}.  Norm conserving Troullier-Martins \cite{troulliermartins} pseudopotentials factorized in the Kleinman-Bylander \cite{kleinmanbylander} form were used. All calculations in this paper employed optimized double $\zeta$ polarized basis sets (DZP), where  two $s$ and three $p$ orbitals for the H valence electron and two $s$, six $p$ and five $d$ orbitals for Si valence electrons were used. The structures were relaxed until the forces on each atom were less than 0.04 eV/\AA. We used a plane wave cutoff of 100 Ry for the grid (used for computing multi-center matrix elements) with $10^{-4}$ for the tolerance of the density matrix  in self consistency steps. We solved the self-consistent Kohn-Sham equations by direct diagonalization of the Hamiltonian and a conventional mixing scheme. The $\Gamma$ point was used to sample the Brillouin zone in all calculations.

Density functional theory in the LDA, or with gradient corrections,  maps the ground state many-electron problem onto a system of non-interacting fermions\cite{kohn}. In principle, the eigenvalues of the resulting single-particle equations are not true excitation energies and the spectral gap between occupied and unoccupied states is well known to be incorrect~\cite{martin}. Nevertheless, the eigenvectors of the problem (the Kohn-Sham orbitals) have been shown to be very similar to quasiparticle states from ``GW" calculations, in which the self-energy is expressed as a product of the single particle Green's function ``G" and the dynamically screened Coulomb interaction ``W" as used in many-body calculations~\cite{hedin}. For Si, C and LiCl, Hybertsen and Louie~\cite{hybertsen} found 99.9\% overlap between ``GW" states and the Kohn-Sham orbitals.  On an empirical level for amorphous materials, there are many indications that it is profitable to interpret the Kohn-Sham orbitals ``literally" for comparisons to experiment~\cite{drabold0,mauri}. This provides some rationale for interpreting the Kohn-Sham orbitals as quasiparticle states, as we shall in some subsequent parts of this paper.

\subsection{Models}

A predictive simulation requires a physically plausible model that represents the topology of the network and yields an accurate description for dynamics of the atoms. In this article we have used two different supercell models: a 223 atom a-Si:H model (Model-I) and a 71 atom a-Si:H model (Model-II). These models are generated from a 64 atom and a 216 atom a-Si models which were generated by Barkema and Mousseau \cite{Mousseau} using an improved version of the Wooten, Winer, and Weaire (WWW) algorithm \cite{www} respectively. 

To create the a-Si:H environment a) We started from a 216 atom a-Si model with two dangling bonds, we removed two silicon atoms resulting in the formation of additional vacancies. All of the vacancies except one are then terminated by placing a H atom at about 1.5~\AA~from the corresponding Si atom. This yield a 223 atom Model-I. b) We started from a defect free 64 atom a-Si model, we removed three silicon atoms resulting in the formation of vacancies. All of the dangling bonds are then terminated by placing a H atom at about 1.5~\AA~from the corresponding Si atom to generate a 71 atom Model-II. We then repeated this supercell surgery at other sites to generate an ensemble of three configurations to obtain some insight into the formation of the structure and its bonding in solid state.  Finally these newly generated structures are well relaxed using conjugate gradient optimization technique. While such a procedure is clearly unphysical, it is worth pointing out that the resulting proton NMR second moments of the clusters created are similar to the broad component of the lineshape observed in experiments\cite{pafdad93}.

\subsection{Excited state dynamics and promotion of carriers}

Defects in an amorphous network may lead to localized electron states in the optical gap or in the band tails. If such a system is exposed to band gap light, it becomes possible for the light to induce transitions from the occupied states to unoccupied states. For the present work we will not concern ourselves with the subtleties of how the EM field introduces the transition, we will simply assume that a photo-induced promotion occurs, by depleting the occupied states of one electron ``forming a hole'' and placing the electron near the bottom of the unoccupied ``conduction'' states. The idea is that a system initially at equilibrium will not be after the procedure: Hellmann-Feynman forces~\cite{hellmann} due to the occupation change will cause structural rearrangements, which may be negligible or dramatic, depending on the flexibility or stability of the network, and the localization of the states. The changes in force will initially be local to the region in which the orbitals are localized, followed by transport of the thermal energy.  In general, it is necessary to investigate photo-structural changes arising from various different initial and final states, though only well localized states near the gap have the potential to induce structural change~\cite{fedders,czech,drabold2,drabold-zhang,drabold-li}. The simulated light excited state is achieved by implementing: a) starting from the well relaxed model, we make the occupation change by adding an additional electron just above the Fermi level, b) we keep the system in this excited state for 10ps (20000 MD steps with time step $\tau$=0.5 fs between each MD steps), and maintain a constant temperature T=300K, c) after 10ps, we put the system back into the ground state and relax to minimize the energy.   The method has been described in additional detail elsewhere\cite{czech}.

\section{Hydrogen Dynamics}
\label{secIII}

We have performed extensive MD simulations of network dynamics of a-Si:H both in an electronic ground state (``light-off") and a simulated light-excited states (``light-on") for the two models, Model-I and Model-II described above. In the next sections, we present a detailed calculation of hydrogen diffusion, its mechanisms and consequences on the structural, electronic and vibrational properties in both electronic ground state and light excited state.

\subsection{Hydrogen motion: Ground State}

To analyze the diffusion mechanism in the ground state we performed a MD simulation for five different temperatures, and tracked the trajectories and bonding information of all the H and Si atoms in the network. In all the cases, the MD simulations show diffusion of hydrogen in the cell and as a consequence, the network exhibits bond break and formation processes. The pattern of diffusion differs for individual H atoms depending upon the geometrical constraints around the diffusing H atom. 

\begin{figure}[h]
\begin{center}
\includegraphics[angle=270, width=0.48\textwidth]{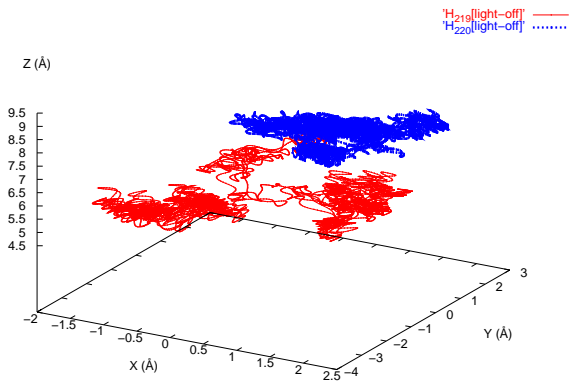}
\end{center}
\caption{\label{fig1}  Trajectory for three different hydrogen atoms (H$_{219}$ and H$_{220}$) in the ground state, which shows the diffusion and trapping of the atom for Model-I model. The total time for the trajectory is 10ps. }
\end{figure}

In order to characterize the trajectories of H diffusion in the ground state, we have selected two diffusive H atoms, (H$_{219}$ and H$_{220}$), and plotted their trajectories at T=300K in Fig.~\ref{fig1}.
The trajectories for both  H$_{219}$ and H$_{220}$ atoms show diffusion in which the H atoms spend time being trapped in a small volume of the cell which is followed by rapid emission to another trapping site. In order to examine how the bond rearrangement takes place in the network while the H atom is diffusing, we tracked each hydrogen atoms and computed its bonding statistics. 

\begin{figure}[h]
\begin{center}
\includegraphics[angle=0, width=0.48\textwidth]{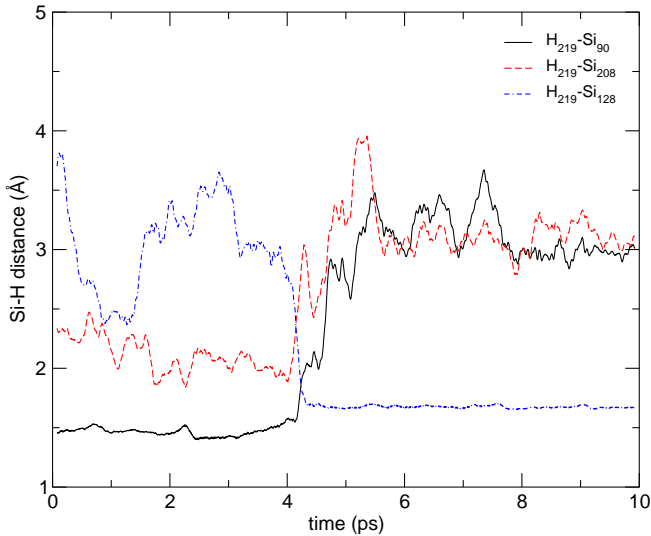}
\end{center}
\caption{\label{fig2} The Si-H bond length between the diffusing H (H$_{219}$) and three different Si atoms, (Si$_{90}$, Si$_{208}$, and Si$_{128}$ as a function of time in the electronic ground state for Model-I. The total time for the trajectory is 10ps. }
\end{figure}

In Fig.~\ref{fig2} we show the Si-H bond length between one of the diffusing H atoms (namely H$_{219}$) and relevant Si atoms (Si$_{90}$ and Si$_{128}$) with which it forms a bond while diffusing and Si$_{208}$. 
As we can see from Fig.~\ref{fig2}, in the first 4ps H$_{219}$ is bonded with Si$_{90}$ with a bond length of 1.5~\AA~and trapped for a while until it breaks and hops to form another bond with Si$_{128}$. In the first $\sim$4 ps, the bond length between H$_{219}$ and Si$_{128}$ fluctuates between 3.8~\AA~and 2.5~\AA.  However, after $\sim$4 ps we observed a swift bond changes in a very short period of time $\sim$0.1 ps when the H$_{219}$ atom comes out of the trapping site and hops to form a bond with Si$_{128}$ and trapped there for a very long time period of $\sim$6 ps. This process of trapping and hopping is typical for the highly diffusive H atoms.

\begin{figure}[h]
\begin{center}
\includegraphics[angle=0, width=0.48\textwidth]{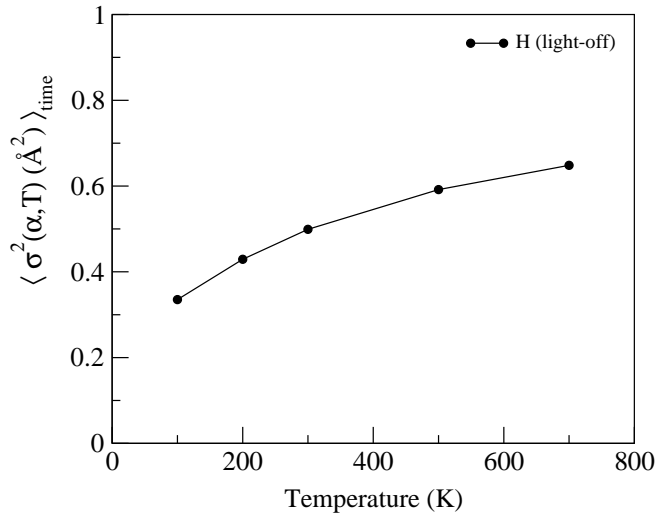}
\end{center}
\caption{\label{fig3} Time average mean square displacement for H as a function of temperature of MD simulation in electronic ground state for Model-II model . }
\end{figure}

To study atomic diffusion we computed the time average mean squared displacement for both H and Si atoms for a given temperature using
\begin{equation}
\label{labeleq1}
\langle \sigma ^2(\alpha, T) \rangle_{\text time} = \frac{1}{N_{\small MD}} \frac{1}{N_ \alpha} \sum_{t=1}^{N_{\small MD}} \sum_{i=1}^{N_\alpha}  |\vec{r}_i{^{\alpha}}(t)-\vec{r}_i{^{\alpha}}(0)|^2,
\end{equation}
where the sum is over particular atomic species $\alpha$ (Si or H), $N_\alpha$ and $\vec{r}_i{^{\alpha}}(t)$ are total number and coordinates of the atomic species $\alpha$ at time $t$ respectively, and $N_{\small MD}$ is the total number of MD steps.  

The time average mean square displacement for Model-II for five different temperatures was calculated using Eq.~(\ref{labeleq1})~for H atoms in the supercell in the electronic ground state (light-off) and it is shown in Fig.~\ref{fig3}. We have observed a strong temperature dependence of H diffusion. This result will help us to compare the diffusion of H in the electronic ground state with the light excited state to be discussed in the next section.

\subsection{Hydrogen diffusion: light excited state}

Similar to the case of electronic ground state, we analyzed the diffusion of H in the light excited state by performing a MD simulation. We tracked the trajectories and bonding statistics of Si and H atoms in the supercell. Our MD simulation in the light excited state show enhanced hydrogen diffusion and consequently increased bond breaking and formation that leads to structural changes in the network. 

\begin{figure}[h]
\begin{center}
\includegraphics[angle=270, width=0.48\textwidth]{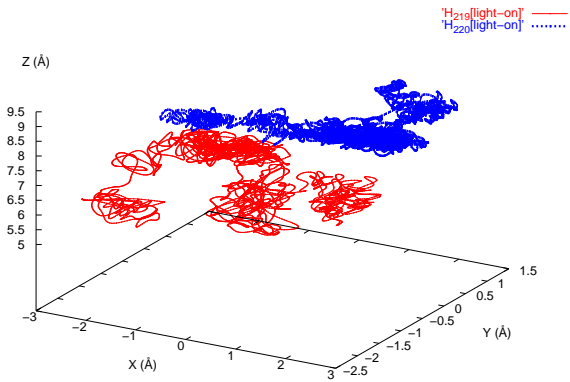}
\end{center}
\caption{\label{fig4}  Trajectory for three different hydrogen atoms (H$_{219}$ and H$_{220}$) which shows the diffusion and trapping of the atom for Model-I in the light excited state. The total time for the trajectory is 10ps. }
\end{figure}

For the purpose of analyzing the difference in the diffusion mechanism of H in the light excited state case as compared with the ground state, we performed similar calculations described in the previous sections for the light excited state case. To see the trajectories of H in the light excited state, we have again selected two diffusive H atoms, (H$_{219}$ and H$_{220}$) from the larger Model-I, and plotted their trajectories in the light excited state in Fig.~\ref{fig4}. The trajectories show the diffusion of H in the presence of different trapping centers, a region where the H atom spends more time before it hops and moves to another trapping site. However, in this case we observed enhanced diffusion and more trapping sites and hopping. These trapping and hopping processes continue until two hydrogens form a bond to a single Si atom to form a metastable SiH$_2$ conformation or until two hydrogens form a bond to (a) two different Si atoms which are bonded to each other, to form (H-Si-Si-H) structure or (b) two different Si atoms which are not bonded but close to each other to form (H-Si Si-H) structure. This is in agreement with a basic event of the H collision model \cite{Branz} and other H-pairing models \cite{kopidakis}.

\begin{figure}[h]
\begin{center}
\includegraphics[angle=0, width=0.48\textwidth]{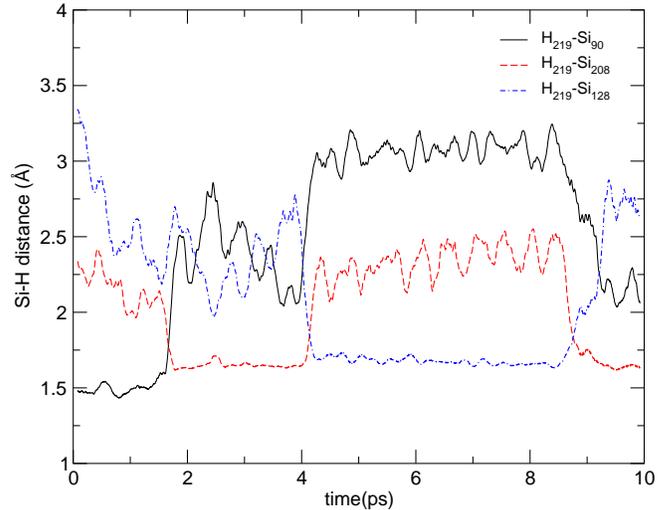}
\end{center}
\caption{\label{fig5} The Si-H bond length between the diffusing H (H$_{219}$) and three different Si atoms (Si$_{90}$, Si$_{128}$, and Si$_{208}$) with which H$_{219}$ forms a bond (one at a time) while it is diffusing as a function of time for Model-I, in the light excited state. The total time for the trajectory is 10ps. . }
\end{figure}

By tracking each H atom, we computed its bonding statistics and examine the bond rearrangements.
In Fig.~\ref{fig5} we show Si-H bond length as a function of time between one of the diffusing H atoms (H$_{219}$) and three other Si atoms (Si$_{90}$, Si$_{128}$, and Si$_{208}$) with which it forms a bond while diffusing in the network. 
The initial trapping time, where H$_{219}$ is bonded with Si$_{90}$, is reduced to $\sim$1.8 ps when the light is on from $\sim$4 ps when the light is off.
This is followed by another trapping site where H$_{219}$ is bonded with Si$_{208}$ for another $\sim$2.1 ps. The H$_{219}$ hops out of the trapping site and forms a bond with Si$_{128}$ and trapped for $\sim$4.3 ps before it finally hops out from the trapping site and forms another bond with Si$_{208}$ where it gets trapped again and form a silicon dihydride (SiH$_2$) structure.
As we can see from Fig.~\ref{fig5}, the pattern of diffusion is quite different from the ground state: In the light excited state case we observed a) more number of trapping sites and less trapping time with frequent hopping, b) enhanced hydrogen diffusion, and c) increasing number of bond rearrangements and newly formed dihydride structural units. 

\begin{figure}[h]
\begin{center}
\includegraphics[angle=0, width=0.48\textwidth]{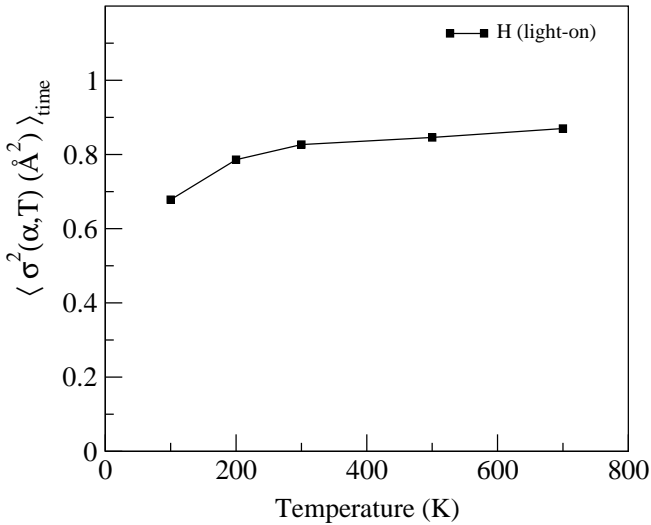}
\end{center}
\caption{\label{fig6} Time average mean square displacement for Has a function of temperature of MD simulation in the light excited for Model-II . }
\end{figure}

The atomic diffusion in the light excited state case has also been examined using the time average mean squared displacement for both H and Si atoms for different temperatures using Eq.~\ref{labeleq1} for both Model-I and Model-II. The results from Model-II is shown in Fig.~\ref{fig6}. 
For all the temperatures considered, our simulation results show enhanced diffusion of Hydrogen for the case when the light is ``on" as compared with the case where the light is ``off". 
Consistent with the work of Isoya~\cite{isoya}, the hopping of H is apparently stimulated by the electron-hole pair. The enhanced diffusive motion of H in the photo excited state relative to the electronic ground state arises from the strong electron-lattice interaction of the amorphous network, and an effect of ``local heating" and subsequent thermal diffusion\cite{drabold2} initially in the spatial volume in which the state is localized. 
The same calculations has also been performed on the larger model Model-I at T=300K in which, the time average mean square displacement for H is 2.66 $\text \AA^2$ for the light excited state and 1.10 $\text \AA^2$ for the electronic ground state. 
These results again show and confirm an enhanced hydrogen diffusion for the case of light excited state. In all the cases no enhanced motion for Si is  observed

\subsection{Consequences of Hydrogen diffusion}

\subsubsection{Formation of dihydride structure}

In the two scenarios that we considered, MD simulation in electronic ground state (light off) and simulated light-excited state (light on) we have observed an important difference. In the light-excited state, in addition to bond rearrangements and enhanced hydrogen diffusion, we have observed a preferential formation of new structure: SiH$_2$, with an average distance of 2.39~\AA~for the pair of hydrogens in the structure, (H-Si-Si-H) and (H-Si Si-H) with H-H separation which ranges from 1.8~\AA~ to 4.5~\AA . However, in the electronic ground state, we have obtained rearrangement of atoms including hydrogen diffusion, without formation of SiH$_2$ structure in the supercell.  The mechanisms for the formation of these structures in the light-excited state follows breaking of H atom from Si-H bond close to the dangling bonds and diffusion to the nearest weakly bonded interstitial sites (or dangling bonds). This mobile H atom then collides (forms a metastable bond) with another Si+DB structure or breaks an Si-Si bond to form another Si-H bond. This is attributed to the fact that the dangling bond site is moving to accommodate the change in force caused by the additional carrier and also because hydrogen is moving through weakly bonded interstitial sites with low activation barrier for diffusion until it is trapped by a defect \cite{santos2}. 

We find that there are two different modes of bond formation  for the mobile hydrogen. The first is when two mobile hydrogen atoms, H$_m$, collide with two Si atoms and form a metastable (H-Si-Si-H) or (H-Si Si-H) structure and the second one is when the mobile hydrogen moves until it encounters a preexisting Si-H+DB structure and makes a bond to form a SiH$_2$ structure.

Consequently, our calculations show two basic ideas for the diffusion of H in the light-excited state: 1) the diffusion of hydrogen doesn't only break a Si-H bond but it also breaks a Si-Si bond and 2) the possibility that two mobile H atoms might form a bond to a single Si atom to form a metastable SiH$_2$ structure in addition to the formation of (H-Si-Si-H) and (H-Si Si-H) structures. 

In the Model-II, the two hydrogens involved in the formation of the SiH$_2$ structure initially were 5.50~\AA~apart and bonded to two different Si atoms (Si-H) which were separated by 4.86~\AA. With thermal simulation in the light excited state, the two hydrogen atoms dissociate from their original Si atoms and becomes mobile until they form the SiH$_2$ structure, in which the H-H distance becomes 2.39~\AA. We have observed similar pattern of H diffusion, bond rearrangements and formation of SiH$_2$ structure near the DB for the other two configurations considered in the simulation. The same phenomenon is observed in the case of Model-I. The two hydrogens involved in the formation of the SiH$_2$ structure initially were 3.29~\AA$~$apart and bonded to two different Si atoms (Si-H) which were separated by 3.92~\AA. With thermal simulation in the light excited state, the two hydrogen atoms dissociate from their original host and becomes mobile until they form the SiH$_2$ structure, in which the H-H distance becomes 2.45~\AA. We have summarized the results that show before and after MD calculations of H-H distance (in SiH$_2$ structure) for Model-I and three different configurations of Model-II in the case of light excited state in Table~\ref{Table2}. 

\begin{table}[htpb]
\caption{\label{Table2} The H-H distance in the SiH$_2$ configurations and the Fermi energy of the system before and after MD simulations in the light excited case.}

\begin{tabular*}{0.48\textwidth}{@{\extracolsep{\fill}}lcc}
\\
\hline\hline
\\
\multicolumn{1}{c}{} &
\multicolumn{2}{c}{H-H distance} \\ 

\multicolumn{1}{c}{Config-} &
 \multicolumn{1}{c}{before MD} & 
 \multicolumn{1}{c}{after MD } \\

 \multicolumn{1}{c}{-urations} &
  \multicolumn{1}{c}{(\AA)} &
  \multicolumn{1}{c}{(\AA)} \\ \cline{1-3}
\\
1 (Model-II)       &  5.50  &  2.39    \\
2 (Model-II)       &  3.79  &  2.36    \\
3 (Model-II)       &  4.52  &  2.36    \\
4 (Model-I)        &  3.29  &  2.45    \\
Average            &            &  2.39    \\ 
\hline \hline
\end{tabular*}
\end{table}

\subsubsection{Change in the electronic properties}
\begin{figure}[htbp]
\begin{center}
\includegraphics[angle=0, width=0.45\textwidth]{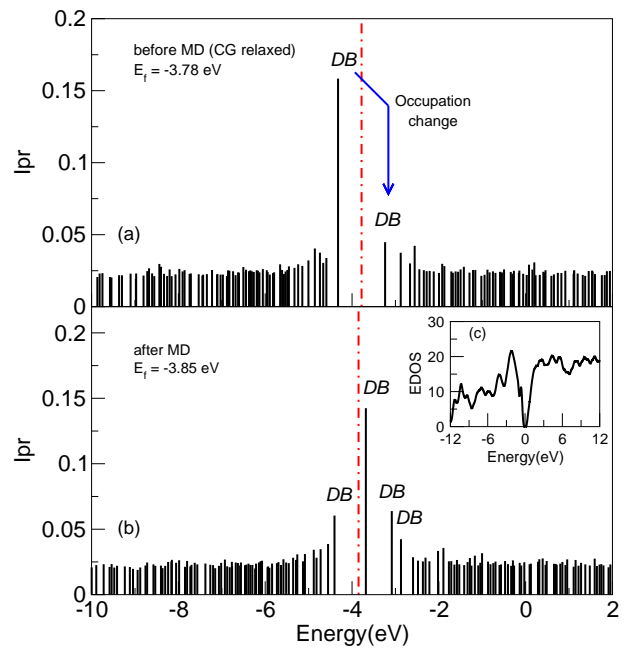}
\caption{\label{fig7} (Color Online) The inverse participation ratio $\mathcal{I}$ of the eigenstates versus the energy eigenvalues, (a) in the relaxed electronic ground state and (b) in the relaxed simulated light-excited state (light excited MD followed by relaxation), with their respective Fermi energy in the first configuration of relaxed Model-II. The inset (c) shows the electron density of states with the Fermi level shifted to zero for the relaxed simulated light-excited state.}
\end{center}
\end{figure}

In order to understand the electron localization we used the inverse participation ratio, $\mathcal{I}$, 
 \begin{equation}
 \mathcal{I} = \sum_{i=1}^{N}[q_i(E)]^2
 \end{equation}
where $q_i(E)$ is the Mulliken charge residing at an atomic site $i$ for an eigenstate with eigenvalue $E$ that satisfies $\sum_{i}^{N}[q_i(E)] = 1$ and $N$ is the total number of atoms in the cell. For an ideally localized state, only one atomic site contributes all the charge and so $\mathcal{I}=1$. For a uniformly extended state, the Mulliken charge contribution per site is uniform and equals $1/N$ and so $\mathcal{I} = 1/N$. Thus, large $\mathcal{I}$ corresponds to localized states. With this measure, we observe a highly localized state near and below the Fermi level and a less localized state near and above the Fermi level. These states, highest occupied molecular orbitals (HOMO) and lowest unoccupied molecular orbitals (LUMO), are centered at the two dangling bonds in the initial configuration of the model. The energy splitting between the HOMO and LUMO states is 1.08 eV. Figure \ref{fig7} (a) shows the Fermi level and $\mathcal{I}$ of these two states and other states as a function of energy eigenvalues in the relaxed electronic ground state.  

\begin{figure}[htbp]
\begin{center}
\includegraphics[angle=0, width=0.45\textwidth]{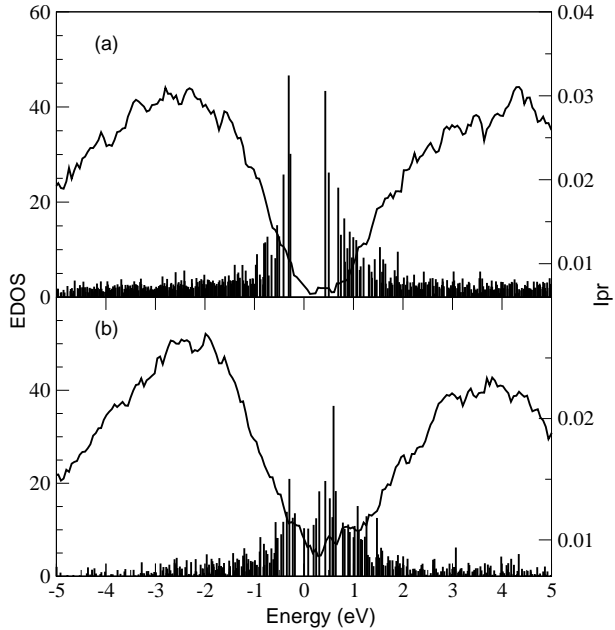}
\caption{\label{fig8} (Color Online) The energy density of states and the inverse participation ratio $\mathcal{I}$ of the eigenstates versus the energy eigenvalues, (a) in the relaxed electronic ground state and (b) in the relaxed simulated light-excited state (light excited MD followed by relaxation), both the electron density of states and the inverse participation ration are plotted with the Fermi level shifted to zero.}
\end{center}
\end{figure}

This picture changes when we excite the system and perform a MD calculation in which we observe enhanced diffusion of hydrogen and subsequent breaking and formation of bonds. Since electron-phonon coupling is large for localized states \cite{attafynn2-n}, the change of occupation causes the forces in the localization volume associated with the DB to change and the system moves to accommodate the changed force. Consequently, the hydrogen atoms close to the DB sites start to move in the vicinity of these defects either to terminate the old DB's or to break a weak Si-Si bond and by doing so, create new DB defects on nearby sites. As shown in Fig. \ref{fig7} (b) we observe the formation a highly localized state and appearance of three less localized states, that correspond to the newly formed defect levels after simulated light-soaking. These processes induce transition of electrons from the top of the occupied states to the low-lying unoccupied states which is reflected in the smaller value of $\mathcal{I}$ for the initial HOMO state and an increase in the $\mathcal{I}$ for the LUMO state. 

The $\mathcal{I}$ of the HOMO state, where the state is initially localized, decreases from 0.158 to 0.060 after photo-excitation, while the $\mathcal{I}$ of the LUMO state increases from 0.045 to 0.142. The splitting energy between the HOMO and LUMO states has also declined to 0.723 eV. The newly formed defects with lower energy splitting between the HOMO and LUMO states suggest a presence of carrier induced bond rearrangements in the supercell. The comparisons for the energy and $\mathcal{I}$ of the system before MD (as relaxed) and after MD is given in Table \ref{Table1}. 

\begin{table}[htbp]
\caption{\label{Table1} The energy and the inverse participation ratio $\mathcal{I}$ of localized states HOMO, LUMO, LUMO+1 and LUMO+2 before and after the MD for Model-II.}

\begin{tabular*}{0.48\textwidth}{@{\extracolsep{\fill}}ccccc}
\hline\hline
\\
\multicolumn{1}{c}{} &
\multicolumn{2}{c}{Eigenvalue} &
\multicolumn{2}{c}{$\mathcal{I}$} \\ \cline{2-5} 
\\
  \multicolumn{1}{c}{} &
  \multicolumn{1}{c}{before MD} & 
  \multicolumn{1}{c}{after MD } &
  \multicolumn{1}{c}{before MD} &
  \multicolumn{1}{c}{after MD} \\
    \multicolumn{1}{c}{} &
    \multicolumn{1}{c}{(eV)} &
    \multicolumn{1}{c}{(eV)} &
    \multicolumn{1}{c}{} &
    \multicolumn{1}{c}{} \\ \cline{1-5}
\\
HOMO      & -4.32  & -4.40  &  0.158 &  0.060 \\
LUMO      & -3.24  & -3.68  &  0.045 &  0.142 \\
LUMO+1    & -2.88  & -3.08  &  0.037 &  0.064 \\
LUMO+2    & -2.66  & -2.87  &  0.030 &  0.042 \\ 
\hline \hline
\end{tabular*}
\end{table}

 In addition,  analysis of the spatial distribution of the configurations shows that the H atoms close to the dangling bonds ($<$ 4.0$~$\AA) are most diffusive and the Si atoms which make most of the bond rearrangements including the Si atom in the SiH$_2$ configurations are close ($<$ 5.50$~$\AA) to the dangling bonds. These show the additional charge carrier induces change in the forces around the dangling bonds and consequently rearranges the atoms around the dangling bond sites and eventually forming an SiH$_2$ structure. On average the newly formed defect sites are 3.80 \AA$~$and 4.70 \AA$~$far away from the two initial defect sites. The newly formed SiH$_2$ structure is (on average) 4.11 \AA$~$away from the initial defect sites.  It is probable that limitations in both length and time scales influence these numbers, but it is clear that the defect creation is {\it not} very local because of the high diffusivity of the H.

The same calculation have been performed on Model-II. In Fig. \ref{fig8} we have plotted both energy density of states and inverse participation ratio as a function of energy in the light excited state case before and after the MD simulation. As can be seen from the figure we obtained more localized states in the middle of the gap which are caused due to an increase in the number of defects upon light excitation. This supports that the diffusion of hydrogen not only forms preferential dihydride structures but also increase the number of defects in agreement with our findings for the smaller cell model of asiH-71.

\subsubsection{Change in the vibrational properties}

For an amorphous solid, the vibrational density of state is a sum of $3N$ (N is the number of atoms) delta functions corresponding to the allowed frequency modes. Starting with the relaxed Model-II subsequent to MD in the light excited state, we computed the vibrational energies (vibrational modes) from the dynamical matrix, which is determined by displacing each atom by 0.02~\AA~in three orthogonal directions and then performing {\it ab initio} force calculations for all the atoms for each displacement to obtain the force constant matrix, and with diagonalization, phonon frequencies and modes.  

\begin{table}[htpb]
\caption{\label{Table3}  Frequency for some of the Si-H vibrational modes of the SiH$_2$ conformation for the first two configurations of the Model-II obtained from our MD simulations and their corresponding experimental values~\cite{cardona,lucovsky,lucovsky2}.}

\begin{tabular*}{0.48\textwidth}{@{\extracolsep{\fill}}cccc}
\\
\hline\hline
\\
\multicolumn{1}{c}{} &
\multicolumn{1}{c}{Rocking} & 
\multicolumn{1}{c}{Scissors}  &
\multicolumn{1}{c}{Stretch} \\ 
  \multicolumn{1}{l}{Configurations} &
  \multicolumn{1}{c}{cm$^{-1}$} &
  \multicolumn{1}{c}{cm$^{-1}$} &
  \multicolumn{1}{c}{cm$^{-1}$} \\ \hline
\\
1      &  629   &  810  &  2025  \\
2      &  625   &  706  &  2047  \\
\\
Experiment                 &  630   &  875  &  2090    \\
 \hline\hline

\end{tabular*}
\end{table}

In our calculations, the VDOS shows H modes of vibrations in the range (600-900) cm$^{-1}$ and also in the range (1800-2100) cm$^{-1}$. We have examined the vibrational modes to pick out those modes arising only from SiH$_2$.  We reproduce the vibrational modes of SiH$_2$ and their corresponding experimental values~\cite{cardona,lucovsky,lucovsky2} in Table \ref{Table3}. The first mode is the rocking mode at 629 cm$^{-1}$ and 625 cm$^{-1}$;  the second is the scissors mode at 810 cm$^{-1}$ and 706 cm$^{-1}$ and the last is the asymmetric stretching mode that occur at 2025 cm$^{-1}$ and 2047 cm$^{-1}$ for the first and second configurations respectively. These results are in good agreement with the IR absorption spectra for the SiH$_2$ structure. The comparison of our results for the vibrational modes of SiH$_2$ with the experiment is summarized in Table \ref{Table3}. The results shown in Table \ref{Table3} are sensitive to the basis sets used in the calculation, in agreement with other work emphasizing the delicacy of H dynamics\cite{vandewalle-n}. 

\section{CONCLUSION}
\label{secIV}

We have presented a direct {\it ab-initio} calculation of network dynamics and diffusion both for the electronic ground state and light-excited state for a-Si:H. We computed the preferential diffusion pathways of hydrogen in the presence of photo-excited carriers. In the light-excited state, we observe enhanced hydrogen diffusion and formation of new silicon dihydride configurations, (H-Si-Si-H), (H-Si Si-H), and SiH$_2$. The two hydrogens in the SiH$_2$ unit show an average proton separation of 2.39~\AA. The results are consistent (a) with the recent NMR experiments and our previous studies, and (b) with the hydrogen collision model of Branz and other paired hydrogen model in the basic diffusion mechanism and formation of dihydride structures. In contrast, simulations in the electronic ground state do not exhibit the tendency to SiH$_2$ formation. Undoubtedly, other H diffusion pathways exist, and the importance of larger simulation length and time scales as well as effects of promotions involving different states (which could include strain defects and floating bonds\cite{dad-fed}) should be undertaken.  For the first time, we show the detailed dynamic pathways that arise from light-induced occupation changes, and provide one explicit example of defect creation and paired H formation.

\section*{ACKNOWLEDGMENTS}
We acknowledge support from the National Science Foundation under NSF-DMR 0310933, 0205858 and the Army Research Office (ARO). We thank Prof. E. A. Schiff for many helpful conversations and suggestions, and Profs. P. A. Fedders and P. C. Taylor for collaboration and discussions.

\end{document}